\documentclass[iop,appendixfloats]{emulateapj}

\usepackage{amssymb,graphics,textcomp,enumerate}
\usepackage[section] {placeins}
\usepackage{mathtools}
\usepackage{chngcntr}
\usepackage{bm}

\newcommand{\ks}{K$_{s}$}

\newcommand{\um}{$\mu$m~}

\shorttitle{Distance to NGC~6822}
\shortauthors{Rich {\it et al.}}

\newcommand{\noprint}[1]{}

\begin{document}


\title{A New Cepheid Distance Measurement and Method for NGC~6822\footnotemark[*]} \footnotetext[*]{This paper includes data gathered with the 6.5 meter Magellan Telescopes located at Las Campanas Observatory, Chile.}


%% \title{Title\footnotemark[1]} \footnotetext[1]{text} (with foootnotetext
%% outside the title), rather that simply a \footnote{}. 

\author{Jeffrey A. Rich,  S. E. Persson, Wendy L. Freedman, Barry F. Madore, Andrew J. Monson, Victoria Scowcroft \& Mark Seibert}
\affiliation{Observatories of the Carnegie Institution of Washington, 813 Santa Barbara St., Pasadena, CA 91101}

\date{\today}

\begin{abstract}
We present a revised distance to the nearby galaxy NGC~6822 using a new multi-band fit to both previously published and new optical, near- and mid-infrared data for Cepheid variables. The new data presented in this study include multi-epoch observations obtained in 3.6~\um and 4.5~\um with the  \emph{Spitzer Space Telescope} taken for the Carnegie Hubble Program. We also present new observations in J, H and \ks~with FourStar on the Magellan Baade telescope at Las Campanas Observatory. We determine mean magnitudes and present new period-luminosity relations in V, I, J, H, \ks, IRAC 3.6~\um and 4.5~\um. In addition to using the multi-band distance moduli to calculate extinction and a true distance, we present a new method for determining an extinction-corrected distance modulus from multi-band data with varying sample sizes. We combine the distance moduli and extinction for individual stars to determine $E(B-V)=0.35\pm0.04$ and a true distance modulus $\mu_{o}=23.38\pm0.02_{stat}\pm0.04_{sys}$.

\end{abstract}

\vfill\eject
\section{Introduction}
As part of an ongoing mission to determine $H_{o}$ to an accuracy of 2\%, the The Carnegie Hubble Program (CHP) has observed several galaxies with known Cepheid variables with the \emph{Spitzer Space Telescope} \citep{Freedman11}. These warm mission data includes multi-epoch observations with the Infrared Array Camera (IRAC) in the 3.6~\um and 4.5~\um passbands (Channels 1 and 2). In this paper we present a Cepheid distance analysis of the local dwarf irregular galaxy NGC~6822 using these new data, as well as new near-IR ground based observations and previously published visible and near-IR data.

Building upon the discovery of variables in NGC~6822 and the first step in the distance scale beyond the Magellanic Clouds by \citet{Hubble25}, the definitive study of the Cepheid population was that of \citet{Kayser67}. She acquired photographic B and V photometry for 13 Cepheids with periods ranging from 10 to 90 days and measured a distance modulus $\mu_{o}=23.75$ mag and a relatively high reddening of $E(B-V)=0.27$ mag. Subsequent re-observation and analysis of these Cepheids proceeded apace, including a near-IR H-band study by \citet{Mcalary83} who found $\mu_{o}=23.47\pm0.11$, with less sensitivity to the high extinction of NGC~6822. More recent multi-band studies have narrowed the range of Cepheid distances and better constrained the average and differential reddening \citep{Gallart96,Pietrzynski04,Gieren06,Feast12}. 

NGC~6822 lies at a low galactic latitude ($b\sim-$18\textdegree) and thus suffers a large line of sight extinction, with derived reddening values varying from $E(B-V)$ of 0.19 to 0.42 mag \citep{Gallart96}. This makes it a prime target for the use of near and mid-IR observations for distance determination. With the accurate calibration of the Cepheid distance ladder using \emph{Spitzer} IRAC data, including well-constrained galactic and Large Magellanic Cloud period-luminosity (PL) relations and distances, we are poised to examine the Cepheids in NGC~6822 at mid-IR wavelengths \citep{Scowcroft11,Freedman11,Monson12}.

A previous study of archival, single-epoch observations of Cepheids in NGC~6822 by \citet{Madore09a} provided motivation and proof-of-concept for the use of IRAC data in measuring the PL relations despite the potential for crowding and high stellar background at mid-IR wavelengths. Recent near-IR studies by \citet{Gieren06} and \citet{Feast12} (henceforth F12) still indicate some uncertainty in the Cepheid distance, even when compared against the previous \emph{Spitzer} data. We discuss this discrepancy and its possible causes.

In this study we follow up on the single epoch data with new 12 epoch \emph{Spitzer} IRAC observations which are used to determine more accurate mid-IR mean magnitudes. We also investigate the calculation of the reddening and true distance modulus with multi-wavelength fits, combining multiple datasets. Our sample has been compiled from the OGLE and Araucaria projects, which have  of V, I, J and \ks~observations of a large number of Cepheids i\citep{Pietrzynski04,Gieren06}. 

When multi-wavelength fitting of a reddening law to apparent distance moduli of Cepheids in external galaxies was first undertaken (\citealt{Freedman85a,Freedman85b,Freedman88a,Freedman91}) its efficacy and novelty largely outweighed its rigorous and self-consistent application.  Distance moduli of individual stars in the galaxy were assumed to be representative, and to be unbiased with respect to one another. As the amount of observations useful for such an analysis grows the method can be refined with an increasing number of passbands to better constrain the line-of-sight extinction and thus the true distance, but for a variety of technical reasons (field of view, detector types, etc.) the stars observed often represent different samplings of the instability strip and individually different line-of-sight extinction. This can lead to a situation where one derives an average reddening and true distance with datasets which may or may not suffer from systematic bias due to different samplings of the instability strip or regions of the galaxy (and thus different reddenings) for each passband used.
\begin{figure*}
\centering
{\includegraphics[width=1.00\textwidth]{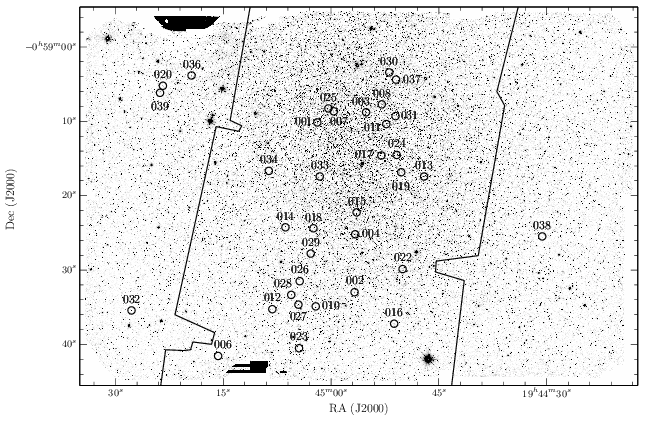}}
\caption{J-Band image of NGC 6822 created from our FourStar data with the overlapping Spitzer IRAC field of view overlaid with black lines. The total Spitzer IRAC field of view extends beyond the northern and southern edges of the FourStar field of view. Cepheids that fall within the FourStar field of view are marked with circles.}
\end{figure*}
To remedy the resulting bias new methodologies need to be explored. One possibility is to keep the basic approach the same, but limit the analysis to a very restricted sample of Cepheids containing (say) only those stars that have data at all of the considered wavelengths, and/or completely dropping from the analysis wavelengths that have the smallest number of observations. The first-suggested, sample restricted approach was recently adopted by F12 in their analysis of Cepheid data for NGC~6822.  While minimizing systematic effects of incoherent samples, the downside of this approach is that large numbers of stars or wavelengths drop out of the analysis and stop contributing to the solution. Some, often hard-won, information is inevitably lost; the trade-off being sample-size reduction (decreased precision) for increased accuracy (decreased bias). 

In this paper, we present a new method for determining extinction and true distance modulus using samples of varying sizes and compositions at multiple wavelengths.  Following the discussion of \citet{Freedman91}, we calculate individual distance moduli for each Cepheid and use the results to generate a true distance modulus gaussian mixture distribution. With our method we hope to bypass some of the possible biases introduced in choosing which Cepheids to use in defining a PL relation. Throughout this paper we adopt a distance to the LMC of $\mu_{o}=18.477\pm0.033$ (\citealt{Freedman12,Monson12}) and the Cepheid PL relations used in \citet{Scowcroft13}, derived by \citet{Fouque07} for the V \& I bands, \citet{Persson04} for the J, H \& K$_{s}$ bands and \citet{Scowcroft11,Monson12} for the IRAC 3.6$\mu$m \& 4.5$\mu$m data.

\section{Sample, Observations and Reduction}
In this paper we use both previously published as well as new observations. To compose our sample of Cepheids, we refer to the work of \citet{Pietrzynski04}, which includes coordinates and the most up-to-date list of Cepheids in NGC~6822, including the re-discovery of Cepheids previously observed with both photographic plates and CCDs (\citealt{Hubble25,Kayser67,Mcalary83,Gallart96,Antonello02}). To simplify matters, we refer to the stars used in our analysis by the unique identifier given in \citet{Pietrzynski04} and adopt the periods determined with their well-sampled data. We restrict our analysis to Cepheids with a periods greater than 6 days, resulting in 39 stars. Their positions are shown in Figure 1, a list of the Cepheids with new observations presented in this work is given in Table 1, and a list of all the Cepheids used for our analysis is given in Table 2. Where observations have already been published, we do not re-reduce or re-photometer the existing data. An example of the phased data points is shown in Figure 2 and the remaining Cepheid data plots are given in the Appendix.

\begin{table}[htpb!]
\vspace{-17pt}
\caption{New Spitzer \& FourStar Photometry} \label{table_mags}
\centering
\begin{tabular}{lcccc}
\hline
\hline
\multicolumn{1}{l}{Name} & \multicolumn{1}{c}{$passband$} & \multicolumn{1}{c}{mag.} & \multicolumn{1}{c}{$\sigma$}  & \multicolumn{1}{c}{MJD}\\
\hline
CEP001 &   [3.6] &  14.44 &  0.05 &   55137.1 \\
CEP001 &   [3.6] &  14.44 &  0.04 &   55151.0 \\
CEP001 &   [3.6] &  14.40 &  0.06 &   55159.5 \\
CEP001 &   [3.6] &  14.62 &  0.04 &   55168.6 \\
CEP001 &   [3.6] &  14.61 &  0.03 &   55345.5 \\
CEP001 &   [3.6] &  14.56 &  0.03 &   55359.3 \\
CEP001 &   [3.6] &  14.48 &  0.05 &   55367.5 \\
CEP001 &   [3.6] &  14.38 &  0.03 &   55376.5 \\
CEP001 &   [3.6] &  14.41 &  0.05 &   55512.4 \\
CEP001 &   [3.6] &  14.55 &  0.05 &   55526.2 \\
CEP001 &   [3.6] &  14.59 &  0.04 &   55535.2 \\
CEP001 &   [3.6] &  14.69 &  0.04 &   55544.7 \\
\hline
\hline
\end{tabular}
\begin{quote}
Multi-epoch photometry of Cepheids measured from our new Spitzer and FourStar data. Table 1 is published in its 
entirety in the electronic edition of ApJ. A portion is shown here for guidance regarding its form and content.
\end{quote}
\end{table}

\begin{figure}
{\includegraphics[width=0.45\textwidth]{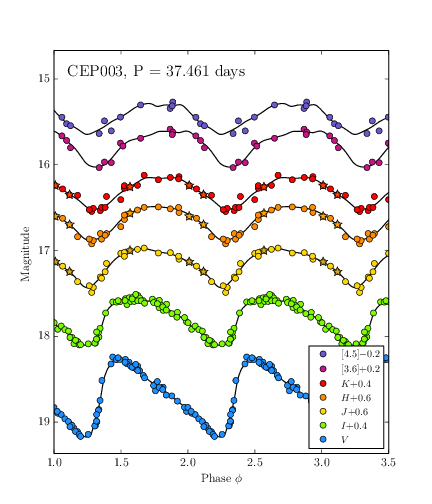}}
\caption{Light curve photometry for a typical Cepheid from our sample. The average magnitude and its error are calculated using the GLOESS fits, shown in this plot as the continuous lines. In cases where the data were not well-sampled, J, H or K$_{s}$ magnitudes are not plotted and were calculated by finding the average value of the data points (see Table 2). The FourStar data, plotted as stars, agree well with those of F12, as discussed in section 2.4. In all cases the J, H and K$_{s}$ fits are dominated by the F12 data. Error bars have been omitted for clarity; they are listed in Table 1.}
\end{figure}

\begin{table*}[htbp!]
\caption{Mean Magnitudes} \label{table_sample}
\centering
\begin{tabular}{lccccccccc}
\hline
\hline
\multicolumn{1}{c}{Cepheid ID} & \multicolumn{1}{c}{P (Days)}  & \multicolumn{1}{c}{V($\sigma$)} & \multicolumn{1}{c}{I($\sigma$)} & \multicolumn{1}{c}{J($\sigma)$}  & \multicolumn{1}{c}{H($\sigma$)} & \multicolumn{1}{c}{\ks($\sigma$)} & \multicolumn{1}{c}{[3.6]($\sigma$)} & \multicolumn{1}{c}{[4.5]($\sigma$)} & \multicolumn{1}{c}{Notes} \\
\hline
%Name    per      vmag         imag       jmag        hmag        kmag        1mag        2mag         note
CEP001 & 123.90 & 17.90(03) & 16.48(03) & 15.61(05) & 15.14(05) & 14.95(05) & 14.56(04) & 14.52(04) &	b	\\
CEP002 & 65.320 & 17.69(03) & 16.44(02) & 15.62(03) & 15.16(02) & 15.00(02) & 14.90(02) & 14.87(03) &	b	\\
CEP003 & 37.461 & 18.69(04) & 17.38(03) & 16.55(03) & 16.04(03) & 15.91(03) & 15.56(04) & 15.63(04) &	a	\\
CEP004 & 34.663 & 18.92(04) & 17.60(03) & 16.77(03) & 16.26(03) & 16.13(02) & 15.88(05) & 15.99(05) &	a	\\
CEP005 & 32.419 & 18.87(03) & 17.68(02) & ...	    & ...       & ...	    & ...      	& ...	    &		\\	    
CEP006 & 31.868 & 18.63(05) & 17.42(03) & 16.58(02) & 16.11(01) & 16.05(01) & 15.82(03) & ...       &	d,e	\\
CEP007 & 30.512 & 19.08(05) & 17.76(03) & 16.84(03) & 16.34(03) & 16.18(04) & 15.85(04) & 15.92(03) &	a	\\
CEP008 & 29.211 & 19.25(04) & 17.95(03) & 16.94(02) & 16.38(01) & 16.36(01) & 16.01(03) & 16.15(03) &	d	\\
CEP009 & 21.120 & 19.45(04) & 18.29(02) & ...	    & ...       & ...	    & ...      	& ...	    &		\\	   
CEP010 & 19.960 & 19.52(04) & 18.27(03) & 17.51(04) & 17.03(04) & 16.90(04) & 16.72(03) & 16.75(02) &	a	\\
CEP011 & 19.887 & 19.90(03) & 18.53(02) & 17.61(03) & 17.04(03) & 16.94(05) & 16.85(03) & 16.86(03) &	a	\\
CEP012 & 19.602 & 19.79(03) & 18.47(02) & 17.68(02) & 17.14(02) & 17.04(02) & 16.85(04) & 16.90(02) &	a	\\
CEP013 & 19.295 & 19.76(04) & 18.61(03) & ...	    & ...       & ...	    & ...      	& ...	    &		\\	   
CEP014 & 18.339 & 19.59(04) & 18.35(03) & 17.59(02) & 17.13(02) & 17.01(03) & 16.75(04) & 16.81(05) &	a	\\
CEP015 & 17.344 & 19.54(04) & 18.38(03) & 17.67(02) & 17.23(02) & 17.11(02) & 16.66(02) & 16.74(02) &	a	\\
CEP016 & 16.960 & 19.91(04) & 18.71(02) & 17.88(03) & 17.38(04) & 17.23(04) & 17.01(04) & 16.88(02) &	a,e	\\
CEP017 & 16.855 & 20.15(03) & 18.79(02) & 17.89(02) & 17.31(02) & 17.23(04) & 16.83(02) & 16.90(02) &	a	\\
CEP018 & 13.872 & 19.92(03) & 18.68(02) & 17.95(02) & 17.51(02) & 17.31(02) & 17.01(02) & 17.07(04) &	a	\\
CEP019 & 11.164 & 19.99(02) & 18.83(01) & 17.99(03) & 17.49(02) & 17.49(02) & 17.00(01) & 17.17(01) &	d	\\
CEP020 & 10.925 & 19.55(03) & 18.69(02) & 17.96(02) & 17.58(01) & 17.60(01) & ...       & ...  	    &	d	\\
CEP021 & 10.783 & 19.91(04) & 18.91(02) & ...	    & ...       & ...	    & ...       & ...	    &		\\	   
CEP022 & 10.277 & 20.56(02) & 19.39(02) & 18.45(08) & 17.98(05) & 18.03(06) & ...       & ...	    &	c	\\
CEP023 & 9.5731 & 20.24(04) & 19.14(02) & ...       & ...       & ...  	    & 17.74(04) & ...  	    &		\\
CEP024 & 9.3664 & 20.49(04) & 19.27(02) & 18.45(09) & 17.98(06) & 18.04(06) & ...       & ...  	    &	c	\\
CEP025 & 8.9367 & 20.94(02) & 19.24(01) & 18.14(02) & 17.42(02) & 17.18(02) & ...       & ...	    &	a,f	\\
CEP026 & 8.4670 & 20.27(03) & 19.14(02) & 18.51(03) & 18.04(01) & 17.90(03) & 17.63(02) & 17.70(01) &	a,g	\\
CEP027 & 8.3912 & 20.58(03) & 19.39(02) & ...       & ...       & ...	    & 17.63(01) & 17.65(02) &		\\
CEP028 & 7.2085 & 19.93(02) & 18.53(01) & 17.52(02) & 16.89(01) & 16.71(02) & ...       & ...  	    &	a,g	\\
CEP029 & 7.1054 & 20.94(03) & 19.75(01) & ...       & ...       & ...	    & ...      	& ...  	    &	g	\\
CEP030 & 6.8842 & 20.98(03) & 19.79(02) & ...       & ...       & ...       & ...      	& ...  	    &		\\
CEP031 & 6.8665 & 20.86(04) & 19.73(02) & ...       & ...       & ...       & ...      	& ...  	    &		\\
CEP032 & 6.7650 & 20.74(02) & 19.75(02) & ...       & ...       & ...       & ...      	& ...  	    &		\\
CEP033 & 6.7356 & 20.97(03) & 19.69(02) & ...       & ...       & ...       & ...      	& ...  	    &		\\
CEP034 & 6.5391 & 20.76(03) & 19.55(02) & ...       & ...       & ...       & ...      	& ...  	    &		\\
CEP035 & 6.2965 & 20.78(03) & 19.67(02) & ...       & ...       & ...       & ...      	& ...  	    &		\\
CEP036 & 6.1457 & 20.58(02) & 19.63(01) & ...       & ...       & ...       & ...      	& ...  	    &		\\
CEP037 & 6.1387 & 21.10(03) & 20.03(02) & ...       & ...       & ...       & ...      	& ...  	    &		\\
CEP038 & 6.0637 & 20.95(04) & 19.84(02) & ...       & ...       & ...       & ...      	& ...  	    &		\\
CEP039 & 6.0011 & 20.72(03) & 19.73(02) & ...       & ...       & ...       & ...      	& ...	    &		\\
\hline
\hline
\end{tabular}
\begin{quote} 
Cepheid ID and periods for all Cepheids with periods greater than 6 days from \citep{Pietrzynski04}\\
\textbf{(a)} F12 and FourStar data used in Gloess fit\\
\textbf{(b)} F12 data only used in Gloess fit, FourStar data excluded due to phase shifting\\
\textbf{(c)} F12 and FourStar data averaged to determine mean magnitude\\
\textbf{(d)} No F12 data available, FourStar data only averaged to determine mean magnitude\\
\textbf{(e)} Spitzer cold data excluded from Gloess fit due to phase shifting\\
\textbf{(f)} Star excluded from further analysis due to crowding (See Figure A.1)\\
\textbf{(g)} Star excluded from further analysis due to deviation from PL relations \citep{Pietrzynski04,Gieren06}
\end{quote}
\end{table*}

%LEAD WITH NEW SPITZER
%published spitzer
%followed by new FourStar
%followed by published IR

\subsection{Warm \emph{Spitzer} Observations}
New mid-infrared observations of NGC~6822 were taken during the Warm \emph{Spitzer} mission as part of a larger survey of Cepheids in external galaxies (PID 61001). Data were taken for twelve separate epochs from November 2009 to December 2010. The data were taken with IRAC at 3.6~\um and 4.5~$\mu$m, with frame times of 30 s. Both IRAC channels were observed simultaneously and the total area covered by the \emph{Spitzer} observations forms a rectangle roughly 0.15\textdegree~by 0.45\textdegree~centered on the galaxy, with the long axis aligned roughly 12 degrees west of north. The field covered in all 12 epochs corresponds to a subset of the rectangle centered on the galaxy approximately 0.11\textdegree~by 0.23\textdegree. Due to the 1.2\arcmin~offset in the FOV of the 3.6~\um channel from the 4.5~\um channel and the  rotation of the spacecraft throughout the year, the number of epochs a star is observed can vary from 4 to 8 in the outer regions of the field to 12 epochs in the main portion (see, e.g., \citealt{Scowcroft13}).

The data reduction and analysis follow the same steps described in detail in \citet{Scowcroft13}. In short, a time-averaged mosaic is created from all 12 epochs to build a deep picture of the field. For the brighter stars, individual epoch mosaics are then used to measure the light curves. For the fainter stars, the deep frame is used for producing an average photometric magnitude. Photometry was carried out with DAOPHOT and ALLFRAME (\citealt{Stetson87,Stetson94}) on both the time-averaged mosaics and the single-epoch mosaics. DAOPHOT was used initially to create a PSF model and object catalog, while final photometry was performed using ALLFRAME. A model PSF for each mosaic was generated using bright, unsaturated, uncrowded stars and allowed to vary in a linear fashion across the field of view. To determine the accuracy of the photometry thousands of artificial stars were added in a grid to each mosaic and the photometry was rerun several times, shifting the grid in a systematic fashion with each subsequent re-run. Artificial star tests resulted in recovery of greater than 95\% of artificial stars down to ${\sim}17.5$ mag at ${\sim}$0.1 mag photometric accuracy. Our \emph{Spitzer} photometry is calibrated using the standard system of \citet{Reach05}.

\subsection{Cold \emph{Spitzer} Data}
We include in this analysis archival, single-epoch observations taken prior to the end of the cryogenic mission by the SINGS survey \citep{Kennicutt03}. Photometry of the Cepheids from those observations was first presented by \citet{Madore09a}, who showed the utility of even single-epoch data for determining mid-IR mean magnitudes for the low amplitude light curves encountered even for luminous Cepheids. Taken together with the new warm \emph{Spitzer} data, the Cepheids in our sample have up to 13 individual epochs. Mean magnitudes should therefore be well determined even for the faintest stars, whose light curves cannot be delineated.

\subsection{FourStar Observations}
NGC~6822 was imaged with the FourStar Infrared Camera \citep{Persson13} on the 6.5~m Baade telescope at Las Campanas Observatory. The J, H and K${_s}$ observations were made on the nights of May 7, 11, 28 2012; the median seeing was 0.45, 0.7, 1.1\arcsec FWHM respectively. The FourStar field of view is 11'$\times$11' (4 detectors with 18\arcsec~gaps and a pixel scale of 0.159\arcsec/pixel) allowing most of NGC~6822 (5'$\times$11') to fit onto two detectors. A 2-position beam-switching technique was used to measure the sky background. In the A-position NGC~6822 is placed on detectors 1 and 2 (aligned N-S) with sky on detectors 3 and 4 (also N-S). In the B-position, NGC~6822 is placed on detectors 3 and 4 with the sky on detectors 1 and 2. A nine-position dither pattern was used at each beam position to mitigate array artifacts and cosmic rays. Final exposure times were 2.1, 2.2 and 3.8 hours at J, H and K${_s}$ respectively. 

Photometry of stars in the resulting 9 FourStar mosaics (three bands for each epoch) was performed using DAOPHOT \citep{Stetson87}.   star tests were carried out in the same manner as described in section 2.1. The artificial stars were accurately recovered brighter than roughly 19th magnitude  in J, H and \ks, with an accuracy in the photometry of approximately 10\% at the faint end increasing to 1\% at the bright end (${\sim}$13th mag), except for the May 7 observations where the accuracy is about a factor of two less due to marginal sampling caused by excellent seeing.

The JH${_s}$ photometry was calibrated to the 2MASS system (using 2MASS stars in the NGC 6822 field itself) for the nights of May 11 and 28. The excellent seeing for the May 7 observations was such that the number of unsaturated stars in the image that match those in the 2MASS catalog was too few and too inaccurate to calculate a calibration. The zero point offset from 2MASS for May 7 was therefore calculated by comparing unsaturated stars from the May 7th catalog to the 2MASS calibrated May 11th catalog.

\subsection{Previously Published near-IR Observations}

Extensive J, H and \ks~observations were done by F12 using the SIRIUS camera on the Infrared Survey Facility Telescope at the South African Astronomical Observatory. These data cover 16 to 19 epochs, thus sampling the Cepheid light curves extremely well. \citet{Gieren06} obtained two to four epochs of J and \ks~observations using the PANIC and SOFI instruments at Las Campanas and La Silla, respectively. F12 noted systematic discrepancies between their measured magnitudes and those of \citet{Gieren06}, though they did not determine a cause. These discrepancies, which amount to mean differences of 0.126$\pm$0.022 mag in J and 0.061$\pm$0.014 mag in \ks, are not only major, but lead to unaccounted for differences in the derived distance moduli. They should therefore be cleared up.

One possible cause of the discrepancies is the photometric calibration applied to each dataset: F12 use 2MASS stars within the field to calibrate their data, as shown in F12. \citet{Gieren06} use separate observations of standard stars to put their data on the UKIRT system, then transform that data to the \citet{Persson04} system, which is virtually identical to the 2MASS system \citep{Carpenter01}. Thus system transformations don't seem relevant to the problem above perhaps a few hundredths of a magnitude. Another possibility lies in the different methods used to calculate mean magnitudes: F12 used Fourier-fits, while \citet{Gieren06} determine a mean magnitude from one to three individual observations using the template fitting method of \citet{Soszynski05}. Phasing problems should not be an issue in the latter method, because the \citet{Gieren06} and \citet{Soszynski05} data were acquired nearly contemporaneously. 

The F12 data sample the light curves much better than do our new FourStar data, and the latter cannot materially improve the F12 mean magnitudes. Nevertheless, the FourStar data are important in deciding which of the two published datasets -- F12 or \citet{Gieren06} -- is correct. We first did GLOESS fits to the F12 light curve data. We then examined the differences between our individual data points and that of the fitted F12 data, on a point-by-point basis. This was done only in those cases where the F12 phase coverage in the neighborhood of the FourStar data points was sufficient to allow a reasonable local GLOESS fit in principle (noise is another matter). The average differences, in the sense FourStar minus F12, are $-0.03 \pm 0.06$ (10, 21), $-0.02 \pm 0.05$ (10, 21), and $0.01 \pm 0.07$ (9,16), in J, H, and \ks, respectively, where the error bars are the dispersions, and the numbers in parentheses are the numbers of stars and comparison points respectively. The rather large scatters in the differences arise mostly from cases in which the F12 data themselves scatter around the fitted light curves, thus leading to uncertain fits. The average differences and their uncertainties ( $\sim$ 0.0.014 mag) are small enough, however, to decisively favor the F12 data over the \citet{Gieren06} data. Our final mean magnitudes were found by inserting the FourStar data directly into the F12 light curves (with no transformations) and using GLOESS to produce the values in Table 2.

\subsection{Previously Published Optical Data}
We do not present any new optical data in this paper, but we do make use of published V and I-band data obtained with the Warsaw 1.3~m telescope at Las Campanas Observatory over 77 nights in 2002. These well-sampled data, taken as part of the OGLE survey and Araucaria projects, were presented by \citet{Pietrzynski04}.

\section{Results}

\begin{figure}[htpb!]
\centering
{\includegraphics[width=0.50\textwidth]{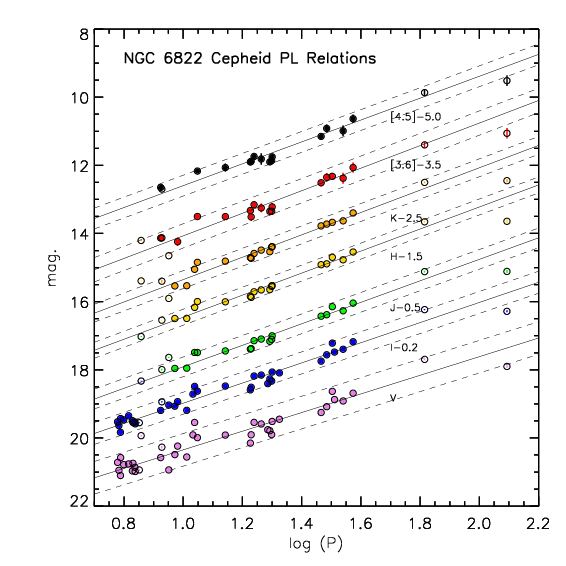}}
\caption{Period-Luminosity relations in all of the bands used in this paper; data are in Table 1. The solid lines represent the unweighted least-squares fits and the dashed lines show the $\pm$2-$\sigma$ deviations. The open symbols are not used in the fitting. Magnitudes are uncorrected for extinction.}
\end{figure}

\subsection{Average Magnitude Determination}
We have re-determined the mean magnitudes in all of the passbands used, including data from the literature. This includes the individual epoch V and I data from \citet{Pietrzynski04}, J, H and \ks~data from F12 (including the FourStar data, and the new multi-epoch IRAC data with a single point added to the light curves where applicable from \citet{Madore09a}. Where possible, we have fitted smooth light curves using the GLOESS program (\citealt{Persson04}). For the V and I-band data we decrease and increase the width of the gaussian smoothing window where the data are well sampled and less well sampled, respectively. The mean magnitudes that we measure for V and I respectively are, on average, fainter than the means calculated by \citet{Pietrzynski04} by $0.03\pm0.02$ and $0.01\pm0.01$ mag. Our J, H and \ks~mean magnitudes are fainter than those found in F12 by $0.01\pm0.02$, $0.00\pm0.01$ and $0.01\pm0.01$ respectively. 

An example of the resulting light curve fits to the data in all available passbands is shown in Figure 2; the remaining fits are given in the Appendix. We omit the FourStar data from the light curves of CEP001 and CEP002 as there appears to be some phase shifting. The \citet{Madore09a} data for CEP006 and CEP016 do not phase well with the new warm \emph{Spitzer} data and have also been omitted from the light curve fit. The mean magnitudes and errors are determined from the light curve fits with GLOESS and are listed in Table 2. The total error in each band is a combination of the systematic (i.e., from fitting) and random photometric errors. As the data are randomly sampled, the photometric error decreases as $1/\sqrt{N}$, where N is the total number of independent observations in each passband.

Of the 22 stars in our sample with J, H and \ks~photometry, four were not measured by F12. For these, the FourStar data are not sufficient to fit light curves. In a study of IC 1613, \citet{Scowcroft13} determined that the shifts in the period and phase over time lead to inaccuracies in template fitting to their FourStar data points. They therefore adopt mean magnitudes and errors by simply averaging the available data points. We do the same for the stars in our sample which lack sufficient near-IR data to fit a light curve: mean magnitudes and their errors are noted in Table 2. Note that the J, H and \ks~magnitudes derived from GLOESS fits are dominated primarily by the F12 data, as our FourStar observations only add two to three data points to the curves (e.g. Fig. 2). Of the 22 stars in our sample with J, H and \ks~magnitudes, only four were not measured by F12.

\begin{table}[htpb!]
\vspace{-17pt}
\caption{PL Relation Slopes and Zero Points} \label{table_sample}
\centering
\begin{tabular}{lccccc}
\hline
\hline
\multicolumn{1}{c}{band} & \multicolumn{1}{c}{Slope (a)} & \multicolumn{1}{c}{Slope (a)} & \multicolumn{1}{c}{Zero-point (b)} & \multicolumn{1}{c}{Zero-point (b)}  &  \multicolumn{1}{c}{N} \\
\multicolumn{1}{c}{}         & \multicolumn{1}{c}{fitted}    & \multicolumn{1}{c}{literature}& \multicolumn{1}{c}{NGC 6822}       & \multicolumn{1}{c}{LMC}  & \multicolumn{1}{c}{} \\
\hline
V       & -2.65 & -2.734 & 23.08 & 14.318 & 34 \\
I       & -2.89 & -2.957 & 22.14 & 13.632 & 34 \\
J       & -3.03 & -3.153 & 21.56 & 13.183 & 17 \\
H       & -3.14 & -3.234 & 21.17 & 12.845 & 17 \\
K$_{s}$ & -3.27 & -3.281 & 21.15 & 12.770 & 17 \\
$[3.6]$ & -3.14 & -3.31  & 20.86 & 12.70 & 16 \\
$[4.5]$ & -2.91 & -3.21  & 20.79 & 12.69 & 14 \\
\hline
\hline
\end{tabular}
\quote{We derive distance moduli by fixing our slope to the literature values in column 2. The references are for V \& I \citealt{Fouque07}; for J, H \& K$_{s}$: \citealt{Persson04}; for $[3.6]$ \& $[4.5]$: \citealt{Scowcroft11,Monson12}. The number of stars used in each fit is given in column 6.}
% Only Cepheids with periods between 6 and 60 days are used.}

% Fix table caption too.

\end{table}

\subsection{Period-Luminosity Relations} 

To determine the PL relation in each passband we use the largest possible sample of mean magnitudes for each color and fit only stars with periods between 6 and 60 days. Following the discussion of \citet{Pietrzynski04} and \citet{Gieren06}, we reject CEP026, CEP028 and CEP029 from the V, I, J, H and \ks~fits due to their distance from the mean PL relation in those studies and CEP025 from the J, H and \ks~fits due to crowding (see Table 2). The PL relations take the form: 

\vspace{-3pt}
\begin{equation*}
M_{\lambda} = a_{\lambda}(\log P - 1.0)+b_{\lambda}
\end{equation*}

An unweighted least-squares fit is performed to each passband; the resulting fit PL relations for each pass-band are shown in Figure 3. The slopes and zero points that we derive are given in Table 3.

\section{Distance Determinations to NGC~6822}

A low galactic latitude leads to a large extinction along the line of sight to NGC~6822, extinction within the galaxy itself, and variable extinction across its face \citep{Fusco12}. It is thus of particular importance to solve for the extinction with high accuracy. Reddening-insensitive Wesenheit magnitudes (\citealt{Madore76,Madore82}) can be calculated using two optical passbands, as in \citet{Pietrzynski04}, resulting in a lower dispersion PL relation and a more accurate extinction-free distance. Alternatively, a distance modulus can be measured for each observed passband and a reddening law fit to the individual distance moduli for each passband, resulting in an extinction-corrected true distance modulus. \citet{Gallart96} present BVRI photometry of six Cepheids outlining the utility of this method and the importance of using an appropriate reddening law \citep{Gieren06,Madore09a,Feast12}. The latter method uses all the available data, and with the addition of  [3.6 $\mu$m] whould give a more robust solution with less extrapolation.

\subsection{Distance and Extinction of the Ensemble}

In this method we use the adopted slope solutions $a_{\lambda}$ given in Table 3 and find $b_{\lambda}$. Offsets of $b_{\lambda}$ with respect to those for the LMC give the apparent distance moduli for each passband, under our assumption that the true (reddening-corrected) distance modulus of the LMC is 18.477$\pm$0.033 mag. Table 4 lists these apparent distance moduli and Figure 4 plots them against $\lambda^{-1}$. The curve is a weighted least-squares fit assuming a \citet{Cardelli89} reddening law with a fixed R$_{v}=3.1$. To account for the varying samples and sets, we estimated the statistical errors in the distance moduli by performing the fit for each band 5000 times while resampling the population with both a random number of objects and a random shift in the measured magnitude within the assigned error bar. The final error is derived from a gaussian fit to the resulting distribution of moduli. We then apply the reddening law of \citet{Cardelli89} to the apparent moduli to determine a true distance modulus $\mu_{o}$ = 23.38$\pm$0.05 and extinction $E(B-V)$ = 0.36$\pm$0.09. We exclude the Spitzer 4.5~\um band from our final $\mu_{o}$ calculation both here and in our subsequent analyses and discussion; in previous studies the 4.5~\um band was found to be contaminated by the CO band-head at 4.6~\um \citep{Freedman11,Scowcroft11,Monson12}.

\begin{table}[htpb!]
\vspace{-17pt}
\caption{Individual Passband Observed Distance Moduli} \label{table_sample}
\centering
\begin{tabular}{lcccccccc}
%\vspace{-10pt}
\hline
\hline
\multicolumn{1}{l}{} & \multicolumn{1}{c}{V} & \multicolumn{1}{c}{I} & \multicolumn{1}{c}{J} & \multicolumn{1}{c}{H} & \multicolumn{1}{c}{\ks} & \multicolumn{1}{c}{[3.6]} & \multicolumn{1}{c}{[4.5]} \\
\hline
$\mu$    & 24.51 & 24.03 & 23.71 & 23.59 & 23.56 & 23.35 & 23.38 \\
$\sigma$ & 0.08 &  0.05 &  0.09 &  0.08 &  0.07 &  0.09 &  0.08 \\
\hline
\hline
\end{tabular}
\end{table}

\begin{figure}[htpb!]
\centering
{\includegraphics[width=0.46\textwidth]{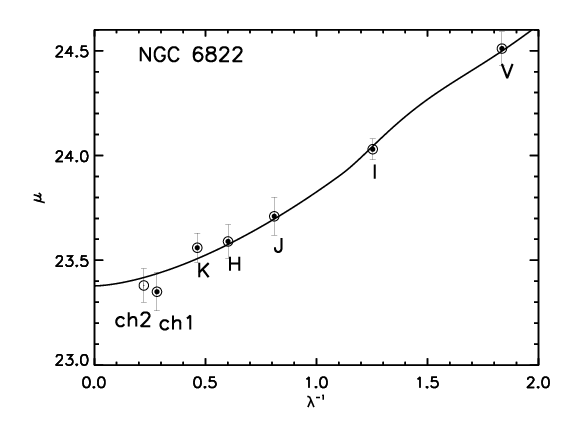}}
\caption{First method for calculating a true distance modulus, $\mu_{o}$ and $E(B-V)$. Data points correspond to the distance modulus from each band plotted as a function of inverse wavelength, $\lambda^{-1}$. A reddening law fit to the moduli (excluding 4.5\um) is plotted as the solid line. The \citet{Cardelli89} reddening law was used with a fixed R$_{v}=3.1$ and $E(B-V)$ and $\mu_{o}$ allowed to vary. The resulting fit provides an extinction-corrected distance modulus of 23.38$\pm$0.05 mag and $E(B-V)$ = 0.36$\pm$0.09.}
\end{figure}

\subsection{Distances and Extinctions of Individual Cepheids}
\begin{figure}[htpb!]
\centering
{\includegraphics[width=0.45\textwidth]{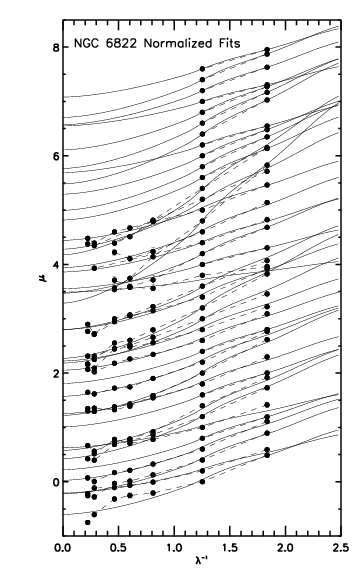}}
\caption{Individual reddening fits applied to each Cepheid, normalized to each star's I-band magnitude and offset from one another by an arbitrary constant amount for plotting purposes. The fits to all 39 Cepheids are shown, increasing in period from CEP039 at the top to CEP001 at the bottom.}
\end{figure}

\begin{table*}[htpb!]
\caption{Derived Individual passband $\mu$ and Fit Values} \label{table_sample}
\centering
\begin{tabular}{lccccccccccc}
\hline
\hline
\multicolumn{1}{l}{Name} & \multicolumn{1}{c}{$\mu_{V}$} & \multicolumn{1}{c}{$\mu_{I}$} & \multicolumn{1}{c}{$\mu_{J}$}  & \multicolumn{1}{c}{$\mu_{H}$}  & \multicolumn{1}{c}{$\mu_{K_{s}}$}  & \multicolumn{1}{c}{$\mu_{3.6}$}  & \multicolumn{1}{c}{$\mu_{4.5}$} & \multicolumn{1}{c}{$\mu_{o}$}  & \multicolumn{1}{c}{$\sigma$}  & \multicolumn{1}{c}{A$_{V}$}  & \multicolumn{1}{c}{$\sigma$} \\
\hline
CEP001 & 25.05 & 24.56 & 24.35 & 24.31 & 24.24 & 23.96 & 23.81 & 23.96 & 0.02 & 1.07 & 0.04 \\
CEP002 & 24.08 & 23.69 & 23.48 & 23.42 & 23.38 & 23.38 & 23.27 & 23.28 & 0.01 & 0.74 & 0.03 \\
CEP003 & 24.41 & 23.92 & 23.65 & 23.53 & 23.49 & 23.24 & 23.26 & 23.30 & 0.02 & 1.09 & 0.04 \\
CEP004 & 24.55 & 24.04 & 23.76 & 23.65 & 23.61 & 23.44 & 23.51 & 23.46 & 0.02 & 1.03 & 0.04 \\
CEP005 & 24.43 & 24.04 & ...   & ...   & ...   & ...   & ...   & 23.46 & 0.08 & 0.96 & 0.10 \\
CEP006 & 24.17 & 23.75 & 23.53 & 23.46 & 23.53 & 23.27 & ...   & 23.38 & 0.01 & 0.67 & 0.04 \\
CEP007 & 24.57 & 24.04 & 23.71 & 23.59 & 23.52 & 23.23 & 23.26 & 23.26 & 0.02 & 1.31 & 0.04 \\
CEP008 & 24.68 & 24.17 & 23.69 & 23.55 & 23.51 & 23.33 & 23.43 & 23.35 & 0.01 & 1.28 & 0.03 \\
CEP009 & 24.50 & 24.10 & ...   & ...   & ...   & ...   & ...   & 23.51 & 0.08 & 0.98 & 0.10 \\
CEP010 & 24.50 & 24.00 & 23.74 & 23.63 & 23.58 & 23.49 & 23.50 & 23.45 & 0.02 & 0.97 & 0.04 \\
CEP011 & 24.88 & 24.26 & 23.85 & 23.65 & 23.58 & 23.61 & 23.61 & 23.48 & 0.02 & 1.33 & 0.03 \\
CEP012 & 24.75 & 24.18 & 23.88 & 23.72 & 23.70 & 23.59 & 23.63 & 23.55 & 0.01 & 1.12 & 0.03 \\
CEP013 & 24.70 & 24.30 & ...   & ...   & ...   & ...   & ...   & 23.70 & 0.08 & 0.98 & 0.11 \\
CEP014 & 24.47 & 23.98 & 23.72 & 23.64 & 23.56 & 23.40 & 23.45 & 23.43 & 0.02 & 0.99 & 0.04 \\
CEP015 & 24.36 & 23.94 & 23.70 & 23.62 & 23.58 & 23.23 & 23.29 & 23.36 & 0.01 & 1.04 & 0.03 \\
CEP016 & 24.69 & 24.24 & 23.90 & 23.75 & 23.69 & 23.55 & 23.41 & 23.42 & 0.01 & 1.33 & 0.03 \\
CEP017 & 24.93 & 24.30 & 23.90 & 23.71 & 23.66 & 23.35 & 23.42 & 23.38 & 0.01 & 1.57 & 0.03 \\
CEP018 & 24.46 & 23.94 & 23.69 & 23.60 & 23.49 & 23.25 & 23.31 & 23.34 & 0.01 & 1.09 & 0.03 \\
CEP019 & 24.28 & 23.81 & 23.44 & 23.28 & 23.21 & 22.94 & 23.11 & 23.01 & 0.01 & 1.32 & 0.02 \\
CEP020 & 23.81 & 23.65 & 23.41 & 23.42 & 23.38 & ...   & ...   & 23.34 & 0.01 & 0.39 & 0.03 \\
CEP021 & 24.16 & 23.85 & ...   & ...   & ...   & ...   & ...   & 23.41 & 0.08 & 0.74 & 0.11 \\
CEP022 & 24.75 & 24.27 & 23.78 & 23.66 & 23.78 & ...   & ...   & 23.54 & 0.03 & 1.21 & 0.05 \\
CEP023 & 24.35 & 23.92 & ...   & ...   & ...   & 23.46 & ...   & 23.38 & 0.04 & 0.92 & 0.06 \\
CEP024 & 24.57 & 24.03 & 23.67 & 23.54 & 23.65 & ...   & ...   & 23.37 & 0.04 & 1.14 & 0.06 \\
CEP025 & 24.96 & 23.94 & 23.28 & 22.88 & 22.70 & ...   & ...   & 22.42 & 0.01 & 2.54 & 0.02 \\
CEP026 & 24.23 & 23.77 & 23.56 & 23.44 & 23.37 & 23.16 & 23.25 & 23.21 & 0.00 & 1.00 & 0.02 \\
CEP027 & 24.53 & 24.01 & ...   & ...   & ...   & 23.16 & 23.19 & 23.09 & 0.01 & 1.48 & 0.02 \\
CEP028 & 23.70 & 22.95 & 22.37 & 22.07 & 21.95 & ...   & ...   & 21.73 & 0.01 & 2.01 & 0.02 \\
CEP029 & 24.69 & 24.16 & ...   & ...   & ...   & ...   & ...   & 23.38 & 0.05 & 1.30 & 0.08 \\
CEP030 & 24.70 & 24.15 & ...   & ...   & ...   & ...   & ...   & 23.35 & 0.07 & 1.34 & 0.09 \\
CEP031 & 24.57 & 24.09 & ...   & ...   & ...   & ...   & ...   & 23.39 & 0.08 & 1.18 & 0.11 \\
CEP032 & 24.44 & 24.09 & ...   & ...   & ...   & ...   & ...   & 23.58 & 0.05 & 0.85 & 0.07 \\
CEP033 & 24.66 & 24.03 & ...   & ...   & ...   & ...   & ...   & 23.11 & 0.07 & 1.53 & 0.09 \\
CEP034 & 24.42 & 23.85 & ...   & ...   & ...   & ...   & ...   & 23.01 & 0.07 & 1.39 & 0.09 \\
CEP035 & 24.39 & 23.92 & ...   & ...   & ...   & ...   & ...   & 23.23 & 0.07 & 1.15 & 0.09 \\
CEP036 & 24.16 & 23.85 & ...   & ...   & ...   & ...   & ...   & 23.40 & 0.04 & 0.75 & 0.05 \\
CEP037 & 24.68 & 24.25 & ...   & ...   & ...   & ...   & ...   & 23.62 & 0.07 & 1.05 & 0.09 \\
CEP038 & 24.52 & 24.04 & ...   & ...   & ...   & ...   & ...   & 23.35 & 0.09 & 1.16 & 0.12 \\
CEP039 & 24.27 & 23.92 & ...   & ...   & ...   & ...   & ...   & 23.40 & 0.06 & 0.87 & 0.08 \\
\hline
\hline
\end{tabular}
\end{table*}

%\subsection{Individual Cepheid Distances}

Although the longer-wavelength \emph{Spitzer} data should provide a more accurate measure of the distance modulus owing to the low extinction at mid-IR wavelengths, the number of Cepheids within the appropriate period range with accurate photometry at 3.6~\um and 4.5~\um is lower than at shorter wavelengths. Not only does this increase the uncertainty in the value of the distance modulus derived with the \emph{Spitzer} data, the \emph{Spitzer} sample does not necessarily match that observed or recovered at other wavelengths, nor will it sample the instability strip sufficiently at all periods. At the shortest periods, for example, Cepheids that fall below the PL relation at visible wavelengths may not be recovered in the \emph{Spitzer} bands where the PSF is relatively large and the crowding effects more serious.

To mitigate the problem of varying sample sizes and selections at different wavelengths, we apply a new method for determining the extinction-corrected distance. Rather than using the PL in each passband and applying a reddening law to find the true modulus for the ensemble, we use all of the available photometry for each individual star to find its true modulus and extinction. The averages over all stars are then the results we seek. The only criterion for this to work is that photometry be available for two or more passbands. 

\begin{enumerate}[(a)]

\item{Take the apparent magnitudes of an individual Cepheid at two or more wavelengths $\lambda_i$ and use the fiducial PL relations to derive its apparent distance moduli $\mu_{\lambda_i}$.} 

\item{Fit a reddening law to the run of apparent distance moduli with inverse wavelength for that single star to determine its individually-estimated true distance modulus, and the statistical error on that modulus.}

\item{Apply steps (a) and (b) for all the stars, with each fit using all the $\mu_{\lambda_i}$ values available. Each fit gives a value of $\mu_o$ at $\lambda^{-1}$ = 0.}

\item{Calculate the weighted mean distance modulus and its statistical error from the individual distance moduli and their uncertainties. The intrinsic dispersion in the instability strip is reintroduced at this step.}

\end{enumerate}

This procedure is in analogy with the ensemble method above, but in this case a $\mu_o$ value returned from the fit is not an actual modulus for that given star, nor is its E(B-V). Each star will appear to have a modulus and reddening that is biased by its position in the instability strip as compared to the mean (ridge-line) solution (See the discussion of this effect by Freedman et al. 1991.) The reason the method works for an ensemble is that we are making the tacit assumption that the instability strip is uniformly filled, and at all periods. This assumption is not, of course, justified for small samples.

For each of the 39 Cepheids in the OGLE sample above a period of 6 days we perform steps (a) \& (b), with the results given in Table 5 and plotted in Figures 5 and 6. The apparent magnitudes for each star are plotted as points, with a least-squares fit deriving the $E(B-V)$ and $\mu_{o}$ for each star using the \citet{Cardelli89} reddening law and R$_{v}$ = 3.1. One advantage of this method is it immediately reveals problems with either the extinction corrections or true moduli. 

Figure 6 shows more clearly how the individual curves converge to a distribution of true distance moduli at $\lambda^{-1}=0$. Each one of the true distance moduli has a corresponding statistical error, which is primarily dominated by the number of bands available to perform the fit

Figure 7 shows a Frequentist approach to deriving the mean true distance modulus and its error. Each Cepheid's true modulus and uncertainty is represented by a unit-area Gaussian having a dispersion equal to the uncertainty in the reddening-law fit to the apparent distance moduli extrapolated to $\lambda^{-1}=0$. Once the mixture distribution is created, the average distance modulus is its median value. Errors are assigned by measuring the cumulative area from this point to a gaussian one $\sigma$, or 34\%, of the area from the mean. The mean and errors measured using this method are $\mu_{o}=23.38\pm0.02$. The final mean $E(B-V)$ from the analogous method is $0.35\pm0.04$.Thus, our two fitting methods give virtually identical results for the true distance modulus: $\mu_o$ = 23.38 $\pm$ 0.05 mag (ensemble method), and 23.38$\pm0.02_{stat}\pm0.04_{sys}$ mag (individual star method), respectively. The individual star method has the advantage of using the data more efficiently: it is not necessary to exclude any sub-samples lacking complete wavelength coverage.

\subsection{Sources of Systematic Error}
Our analysis builds on previously published work done as part of the CHP and as such our analysis is subject to the same sources of systematic error as those studies. \citet{Freedman12} and \citet{Monson12} in particular provide an excellent summary and discussion of the systematic errors associated with the Spitzer data, including photometric zero-points and crowding. Those studies focus on the more nearby Magellanic clouds, so we investigate the effect of crowding on our photometry with artificial stars, following the methodology of \citet{Scowcroft13}. The artificial star tests provide a useful measure of the photometric error (\S\S2.1, 2.3), but we do not find a systematic offset due to crowding in the measured artificial star magnitudes for neither the FourStar nor the Spitzer data we present here, consistent with the study of IC 1613 by \citet{Scowcroft13}. The larger source of systematic error in our photometry is the photometric zero point. We measure an error of 0.01 mag in our FourStar data, and adopt the zero point error of 0.016 mag for the Spitzer data \citep{Reach05}.

The largest source of systematic error in our analysis is the uncertainty in the zero-point of the Cepheid PL relation, which is defined in our work through the adoption of a true distance to the LMC. We adopt the \citet{Freedman12} value of $\mu_o$ = 18.477 mag with a systematic uncertainty of 0.033 mag, as determined with Cepheids and mid-IR data, consistent with our analysis. The value we use is consistent with measurements made with other distance determination methods (see \citealt{Walker12} for a summary), including a recent eclipsing binary measurement by \citet{Pietrzynski13}. When added in quadrature, the sum of our systematic errors results in a total systematic error of 0.04 mag.

\begin{figure}[htpb!]
\centering
{\includegraphics[width=0.45\textwidth]{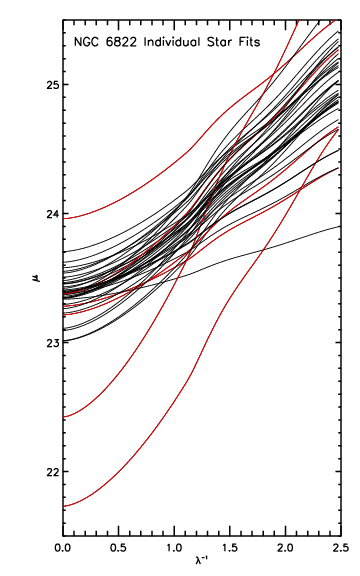}}
\caption{ Same as Figure 5, but without normalization applied and individual data points removed.. All 39 Cepheids are plotted, each with a smooth curve that follows the adopted reddening law. Cepheid fits excluded from the final distance calculation are plotted in red; these include the steep outliers and those discussed in the text. The lines of reddening intercept $\lambda^{-1}=0$ at the true modulus.}
\end{figure}

\begin{figure*}[htpb!]
\centering
{\includegraphics[width=1.00\textwidth]{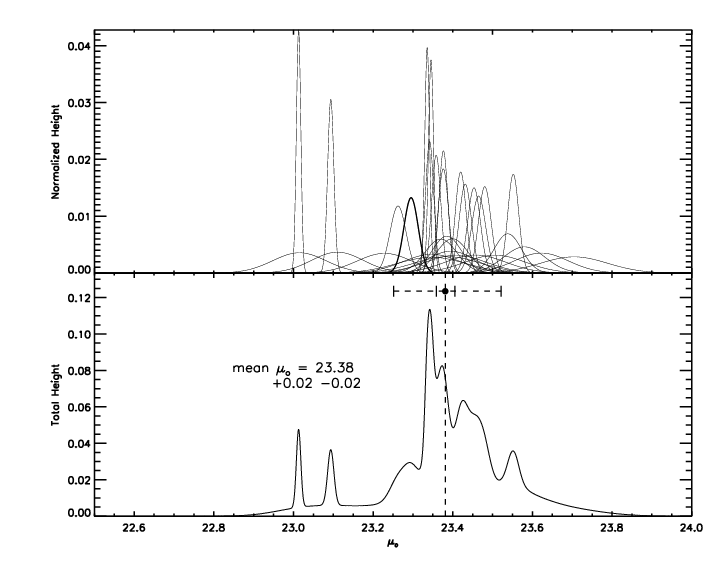}}
\caption{Mixture distribution of distance moduli in Table 5. Top panel: Normal distributions with centers defined by $\mu_{o}$ and width ($\sigma$) corresponding to the error in $\mu_{o}$. Bottom: sum of the normal distributions in the top panel. The dashed vertical line corresponds to the mean, also plotted as a point above the distribution. The widest error bars correspond to the 1$\sigma$, or 34.1\% of the total area starting from the center outward; the narrower error bars correspond to the the error in the mean divided by the square root of the number of stars used in the fit, i.e., $\sqrt{33}$. }
\end{figure*}

\begin{table*}
\caption{True Distance Moduli from the Literature} \label{table_sample}
\centering
\begin{tabular}{lccccl}
\hline
\hline
\multicolumn{1}{l}{Type} & \multicolumn{1}{c}{$\mu_{o}$} & \multicolumn{1}{c}{$\sigma_{stat}$} & \multicolumn{1}{c}{$\sigma_{sys}$} &  \multicolumn{1}{c}{$\sigma_{tot}$} & \multicolumn{1}{l}{Reference}  \\ % added _stat, & sigma_sys
\hline
Cepheid ($VRIJHK_{s}[3.6]$) & 23.38 & 0.02 & 0.03 & ---  & This paper \\
Cepheid ($BVRI$)         	& 23.49 & 0.08 & ---  & ---  & \citet{Gallart96} \\ %E(B-V) error of 0.03?
Cepheid ($W_{V-I}$)      	& 23.34 & 0.04 & 0.05 & ---  & \citet{Pietrzynski04} \\
Cepheid ($\bm{V}$)          & 23.29 & 0.05 & ---  & ---  & \citet{Pietrzynski04} \\ %changed W_V to V
Cepheid ($\bm{I}$)          & 23.32 & 0.05 & ---  & ---  & \citet{Pietrzynski04} \\ %changed W_I to I
Cepheid ($JK_{s}$)		 	& 23.31 & 0.02 & 0.06 & ---  & \citet{Gieren06} \\
Cepheid ($VIJHK_{s}$)    	& 23.43 & 0.02 & ---  & ---  & \citet{Feast12} \\ 
RR-Lyrae                 	& 23.36 & ---  & ---  & 0.17 & \citet{Clementini03} \\ %systematic rolled into total error?
TRGB                     	& 23.46 & ---  & ---  & 0.10 & \citet{Lee93} \\ %Not sure?
TRGB                     	& 23.20 & 0.05 & 0.2  & ---  & \citet{Wyder03} \\
TRGB                     	& 23.53 & 0.05 & ---  & ---  & \citet{Fusco12} \\
Red Clump                	& 23.34 & 0.10 & 0.2  & ---  & \citet{Wyder03} \\
CMD~(Field 1)            	& 23.38 & 0.06 & ---  & 0.06 & \citet{Mcquinn10} \\
CMD~(Field 2)            	& 23.39 & ---  & ---  & 0.06 & \citet{Mcquinn10} \\
CMD~(Field 3)            	& 23.45 & ---  & ---  & 0.06 & \citet{Mcquinn10} \\
CMD                      	& 23.40 & ---  & ---  & ---  & \citet{Fusco12} \\
Mira                     	& 23.56 & 0.03 & ---  & ---  & \citet{Whitelock13}\\
\hline
\hline
\end{tabular}
\begin{quote}
\centering
The errors given above correspond to the statistical, systematic or total error as defined by each reference.
\end{quote}
\end{table*}

\section{Comparison with Previously Measured Distances}
The distance moduli found using the two related methods agree well, which should be the case since they use similar datasets and assumptions about PL slopes. The new method based on individual stars, however, has the advantage of using the data more efficiently: it is not necessary to exclude any sub-samples lacking complete wavelength coverage.

Table 6 compiles distance moduli, reddenings, and their associated errors reported in the literature. 

\subsection{Cepheid Distance Comparison}
The study with the largest sample of Cepheids by far is the work by \citet{Pietrzynski04} as part of the Araucaria project. Their V \& I observations combined with A$_{V}=1.66$ mag and A$_{I}$=0.71 mag (using $E(B-V)=0.36$ mag from \citealt{Mcalary83} and the reddening maps of \citealt{Schlegel98}) yields extinction-corrected distance moduli $\mu_{oV}=23.29$ mag and $\mu_{oI}=23.32$ mag which they compare with their calculated Wesenheit W$_{VI}$ value of $\mu_{W}=23.338$ mag. Interestingly, when we restrict our new method to only the V and I bands, we obtain a mean $E(B-V)=0.37$ mag and a modulus of $\mu_{o}=23.29$ mag which, accounting for the difference in $\mu_{LMC}$ used, agrees with \citet{Pietrzynski04}. When applying the the V \& I-band PL relation slopes derived by the OGLE project (\citealt{Udalski99,Udalski00}), we find the same distance modulus $\mu_{o}=23.38$, with a slightly higher mean $E(B-V)=0.38$. %changed "OGLE" to "Araucaria" in the first sentence, added note about OGLE slopes

The addition of multiple bands helps constrain the distance measured. BVRI photometry of 6 Cepheids by \citet{Gallart96} yielded a higher value of $\mu=23.49\pm0.08$ mag, though the reddening found in that work was a lower $E(B-V)$=0.24 mag. The addition of even a single near or mid-IR band helps to further constrain the extinction and correspondingly the measured distance modulus. The work by \citet{Gieren06} as part of the Araucaria project combined J and \ks~observations with the OGLE V and I data and found a total reddening of $E(B-V)$=0.356 mag and distance $\mu_{o}=23.312$ mag. When we restrict our new method to V, I, J and \ks~only we find a higher $\mu_{o}=23.39$ mag, though this is likely due to the discrepancy between the J and \ks~values used in our work: our adopted mean $\mu_{J}$ and $\mu_{K_{s}}$ are 0.14 and 0.10 mag fainter, respectively.

As our near-IR light curves are dominated by data from F12, the near-IR magnitudes we measure are consistent with those published in that study. With a few exceptions, our individual observations phase well within the light curves observed by F12. As such, the near-IR distances we derive are essentially equal to the apparent moduli in F12. They use a consistent set of Cepheids across all bands, and also re-fit the distance moduli in all bands using their own PL relations. Despite the agreement between our near-IR observations, the $\mu_{V}$ and $\mu_{I}$ F12 find are 0.32 and 0.14 mag lower than the values we derive, which leads to a correspondingly lower $E(B-V)=0.215$ mag in their work. This discrepancy is due to the difference in zero point and slope between their derived PL relation and the zero point and slopes we apply \citep{Fouque07}. This difference accounts for the higher $\mu_{VIJHK}=23.43$ mag in their work compared to our $\mu_{VIJHK}=23.36$.

\subsection{TRGB Distance Comparison}
In addition to the recent and substantial work on Cepheid distances, there are also a number of tip of the red giant branch (TRGB) distance measurements available for comparison to our work. The TRGB traces stars older than Cepheids and provides an independent measure of the distance.  On average, TRGB measurements taken from the literature produce a larger distance than those measured with Cepheids, including our measurement.

The most recent study calculating a TRGB distance to NGC~6822 is \citet{Fusco12}. They perform a new photometric analysis of archival \emph{HST ACS} observations, obtaining color-magnitude diagrams (CMD) for three fields. They determine a reddening using the similar CMD of the nearly extinction-free IC 1613, finding $E(B-V)=0.30\pm0.032$ mag in their outer fields and $E(B-V)=0.37\pm0.02$ mag in their inner field. The higher extinction in the central field calculated by \citet{Fusco12} is consistent with the average reddening we calculate using our Cepheids, which have a distance from the center consistent with the location of their central field.

\citet{Fusco12} use a new theoretical calibration to determine a TRGB distance of $\mu_{o}=23.54\pm0.05$ mag, which is 0.16 mag larger than our distance, but consistent with the mean of other TRGB distance measurements. Although the values do not overlap within the statistical error, \citet{Fusco12} note that their value is consistent with the Cepheid distance of F12 when the offset in the LMC distance is taken into account. This does not, however, account for the discrepancy between our value and that of Fusco et al. given the slightly larger difference in distance modulus and decreased uncertainty in $\mu_{LMC}$ \citep{Freedman11,Monson12}. The discrepancy may be due to metallicity. Fusco et al. note the tip of the red giant branch varies by 0.08 magnitudes over [Fe/H] from -0.8 to -2.0. \citet{Lee93}, for instance, find $\mu_{TRGB}=23.46\pm0.1$ and [Fe/H]$=-1.8$; while Fusco et al.  note in their discussion of metallicity effects that a value of $\mu_{o}=23.40$ would imply a much higher [Fe/H]$=-0.8$.

\subsection{Other Methods}
Some other distance measurements have been explored in the literature and can be compared with our Cepheid distance calculation. \citet{Clementini03} examine a sample of short period variables in NGC~6822 and determine a distance using RR-Lyrae variables, which trace older field stars and provide an alternative pathway to distance calibration. They find a distance of $\mu_{o}=23.36\pm0.17$, which, given the error, is consistent with both our estimate and the TRGB distance given in \citet{Fusco12}.

\citet{Mcquinn10} use \emph{HST WFPC2} data and a CMD fitting routine to determine distance moduli of $\mu_{o}=23.38, 23.39$ and $23.445$ mag in three different fields. They also fit a corresponding A$_{V}=1.1, 1.1$ and $0.7$ mag, consistent with our results. \citet{Wyder03} also examines \emph{HST WFPC2} data and derive a CMD in several fields, and find a distance using the red clump of $\mu_{o}=23.34\pm0.1$ mag. The TRGB distance derived by Wyder is much lower, at $\mu_{o}=23.20\pm0.05(random)\pm0.2(systematic)$ mag, which the author attributes to a large systematic uncertainty due to chemical enrichment. This is in contrast with \citet{Fusco12}, whose $\mu_{cmd}=23.40$ is 0.14 mag less than their $\mu_{TRGB}$.

Finally, \citet{Whitelock13} use Mira variables to derive a distance of $\mu_{o}=23.56\pm0.03$ mag. Even accounting for the 0.03 mag difference in the $\mu_{LMC}$ used (18.50 vs. our 18.47) this value deviates significantly from our measurement. \citet{Whitelock13} adopt a lower extinction value of A$_{V}=0.77$ mag from \citet{Clementini03} and extinction values from \citet{Schlegel98}. Even with their near-IR observations, when compared with our adopted extinction this can lead to a difference of 0.09 magnitudes at J and 0.05 magnitudes at K which begins to account for the difference.

\section{Conclusions}
In this study we presented a new distance measurement to the nearby dwarf starburst NGC~6822 derived using multiple passband observations. In addition to previously published data we utilized new, previously unpublished mid-infrared \emph{Spitzer} data and new near-IR FourStar observations. Using both old and new multi-epoch data, we calculated new mean magnitudes for 39 cepheids in the V, I, J, H, \ks, 3.6~\um and 4.5~\um bands.

In addition to calculating distance moduli for each band and calculating an average line-of-sight extinction and true distance, we introduce a new method for dealing with different samples of available Cepheids at different wavelengths. We use the mean magnitudes for each individual star to fit a distance modulus $\mu$ to that star and corresponding extinction. We then use the resulting individual moduli to create a mixture distribution from which we derive a true, extinction-corrected distance modulus of $\mu_{o}=23.38\pm0.02$ mag. 

We compare our results with both previous distance measurements using Cepheid variables as well as other distance measures. Our value agrees broadly with other distances calculated in the literature, though our result shows a discrepancy compared to the newest calculation of a TRGB distance to NGC~6822 by \citet{Fusco12}. 

\begin{acknowledgements}
This research has made use of the NASA/IPAC Extragalactic Database (NED) which is operated by  the Jet Propulsion Laboratory, California Institute of Technology, under contract with the National  Aeronautics and Space Administration. This research has also made use of NASA's Astrophysics Data System, and of SAOImage DS9 \citep{joye03}, developed by the Smithsonian Astrophysical Observatory.

\end{acknowledgements}

\bibliographystyle{apj}
\bibliography{ms}

\pagebreak

\appendix
\setcounter{figure}{0} \renewcommand{\thefigure}{A.\arabic{figure}} 
\setcounter{table}{0} \renewcommand{\thetable}{A.\arabic{table}} 
This appendix contains supplemental information about the objects and analysis used in this paper. Table A1 lists the coordinates of the cepheids used in this paper, adopted from \citet{Pietrzynski04}, adjusted slightly to account for an apparent offset of approximately 1.4\arcsec west and 0.8\arcsec north. Figure A.2 shows postage-stamp images of Cepheids with FourStar data used in our analysis. Figs. A.2 through A.11 show the light curves fit to the individual passbands using GLOESS, as shown first in Figure 2.
\\
\begin{figure*}
\centering
\includegraphics[width=1.00\textwidth]{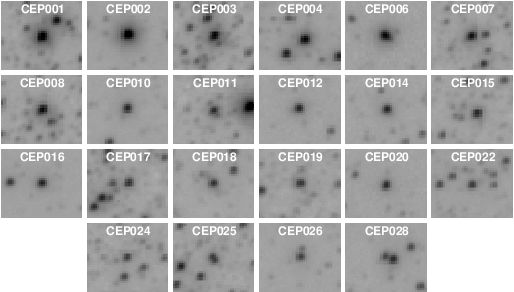}
\caption{Postage stamps of the Cepheids with near-IR magnitudes measured in our analysis. The images are from our May 7, 2012 FourStar observations for which we had exceptionally good seeing.}
\end{figure*}

\begin{table}[htpb!]
\caption{Cepheid Coordinates} \label{table_sample}
\centering
\begin{tabular}{crrcrr}
\hline
\hline
\multicolumn{1}{l}{Designation} & \multicolumn{1}{c}{$RA (J2000)$} & \multicolumn{1}{c}{$Dec (J2000)$} & \multicolumn{1}{l}{Designation} & \multicolumn{1}{c}{$RA (J2000)$} & \multicolumn{1}{c}{$Dec (J2000)$} \\
\hline
CEP001 &        19:45:01.92 & -14:47:32.3 & CEP021 & 19:45:46.53 & -14:56:41.6 \\
CEP002 &	19:44:56.72 & -14:53:14.9 & CEP022 & 19:44:50.06 & -14:52:28.2 \\
CEP003 &	19:44:55.13 & -14:47:11.9 & CEP023 & 19:45:04.46 & -14:55:07.7 \\
CEP004 &	19:44:56.66 & -14:51:17.7 & CEP024 & 19:44:50.84 & -14:48:36.5 \\
CEP005 &	19:45:24.95 & -14:58:44.0 & CEP025 & 19:45:00.37 & -14:47:04.0 \\
CEP006 &	19:45:15.71 & -14:55:23.4 & CEP026 & 19:45:04.37 & -14:52:52.7 \\
CEP007 &	19:44:59.62 & -14:47:10.0 & CEP027 & 19:45:04.55 & -14:53:39.8 \\
CEP008 &	19:44:52.95 & -14:46:55.8 & CEP028 & 19:45:05.57 & -14:53:20.1 \\
CEP009 &	19:45:43.20 & -14:45:42.3 & CEP029 & 19:45:02.84 & -14:51:56.5 \\
CEP010 & 	19:45:02.13 & -14:53:43.4 & CEP030 & 19:44:51.90 & -14:45:51.4 \\
CEP011 & 	19:44:52.32 & -14:47:35.6 & CEP031 & 19:44:51.02 & -14:47:19.8 \\
CEP012 & 	19:45:08.17 & -14:53:49.1 & CEP032 & 19:45:27.76 & -14:53:51.6 \\
CEP013 & 	19:44:47.08 & -14:49:21.1 & CEP033 & 19:45:01.53 & -14:49:21.1 \\
CEP014 & 	19:45:06.35 & -14:51:04.0 & CEP034 & 19:45:08.67 & -14:49:09.9 \\
CEP015 & 	19:44:56.46 & -14:50:33.5 & CEP035 & 19:45:01.49 & -14:43:09.3 \\
CEP016 & 	19:44:51.22 & -14:54:18.3 & CEP036 & 19:45:19.40 & -14:45:57.6 \\
CEP017 & 	19:44:52.99 & -14:48:39.0 & CEP037 & 19:44:50.96 & -14:46:05.2 \\
CEP018 & 	19:45:02.45 & -14:51:06.0 & CEP038 & 19:44:30.61 & -14:51:21.3 \\
CEP019 & 	19:44:50.25 & -14:49:12.8 & CEP039 & 19:45:23.78 & -14:46:32.5 \\
CEP020 & 	19:45:23.38 & -14:46:18.2 & ...    & ...~~~~~~~  & ...~~~~~~~  \\
\hline
\hline
\end{tabular}
\end{table}

\begin{figure}
\centering
\includegraphics[width=0.45\textwidth]{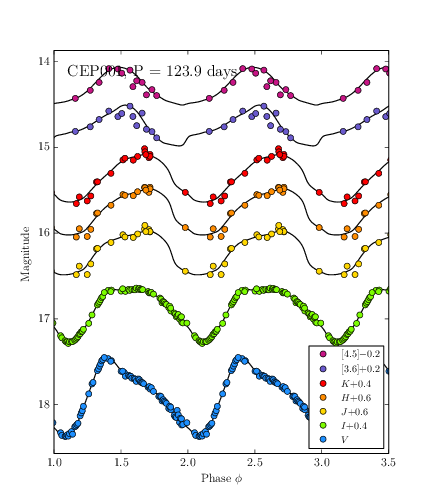}
\includegraphics[width=0.45\textwidth]{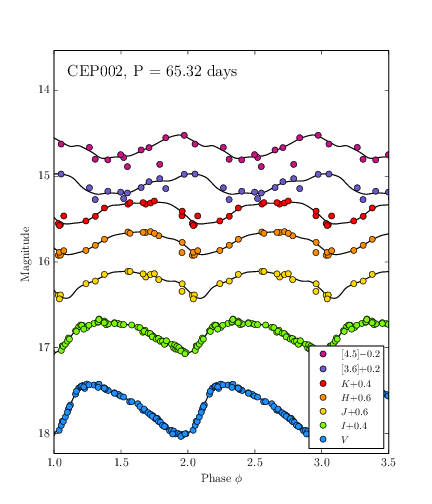}
\includegraphics[width=0.45\textwidth]{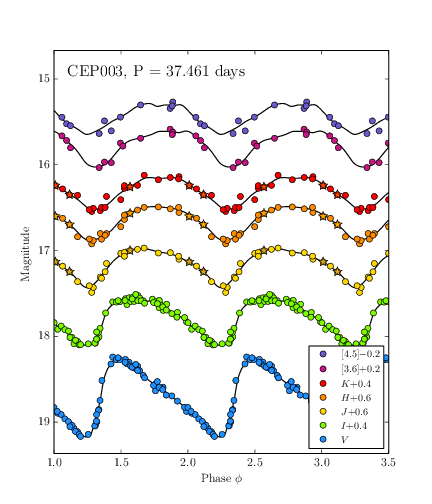}
\includegraphics[width=0.45\textwidth]{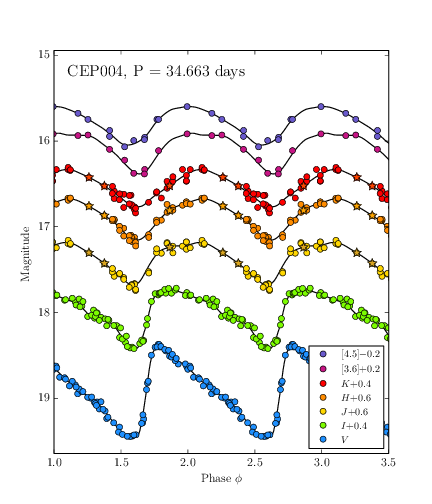}
\caption{Light curves for the GLOESS fits to four Cepheids (CEP001, CEP002, CEP003, CEP004), as in Fig. 2. The average magnitude and its error are calculated using the GLOESS fits, shown in this plot as the continuous lines. In cases where the data were not well-sampled, J, H or K$_{s}$ magnitudes are not plotted and were calculated by finding the average value of the data points (see Table 2). The FourStar data, plotted as stars, agree with those of F12, as discussed in section 2.4. The J, H and K$_{s}$ fits are dominated by the F12 data.}
\end{figure}

\begin{figure}
\centering
\includegraphics[width=0.45\textwidth]{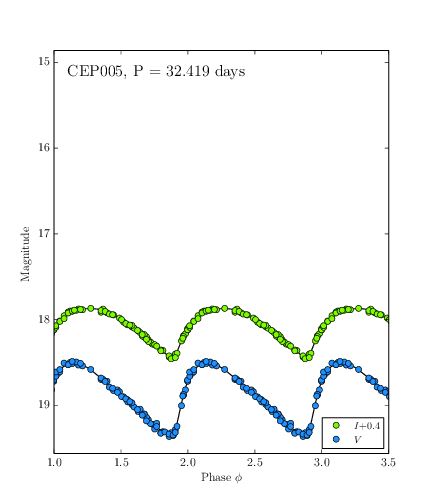}
\includegraphics[width=0.45\textwidth]{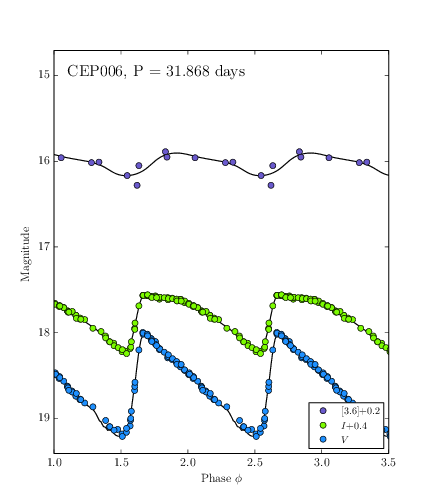}
\includegraphics[width=0.45\textwidth]{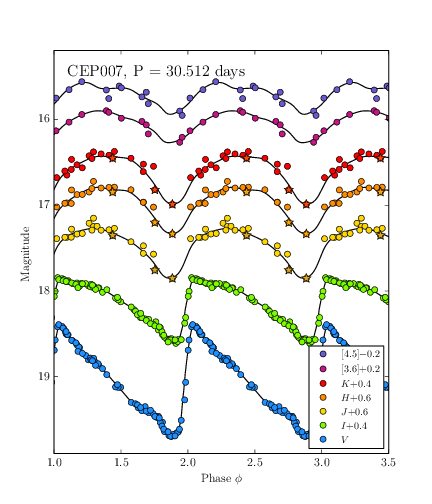}
\includegraphics[width=0.45\textwidth]{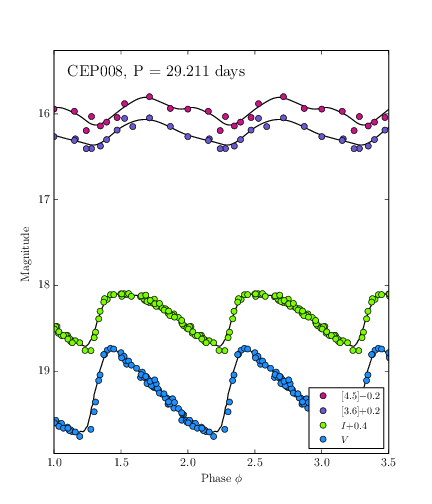}
\caption{Same as Figure A2, but for CEP005, CEP006, CEP007 and CEP008.}
\end{figure}

\begin{figure}
\centering
\includegraphics[width=0.45\textwidth]{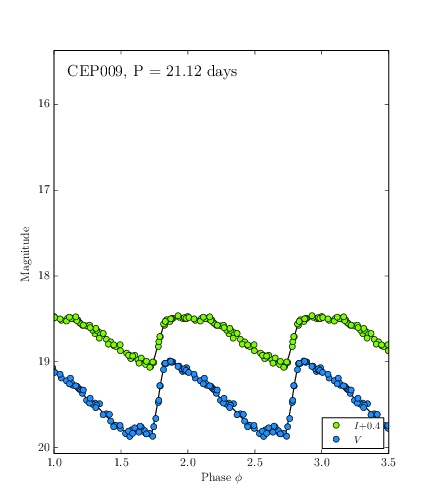}
\includegraphics[width=0.45\textwidth]{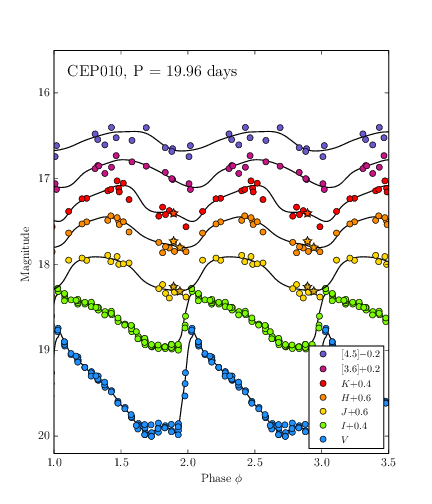}
\includegraphics[width=0.45\textwidth]{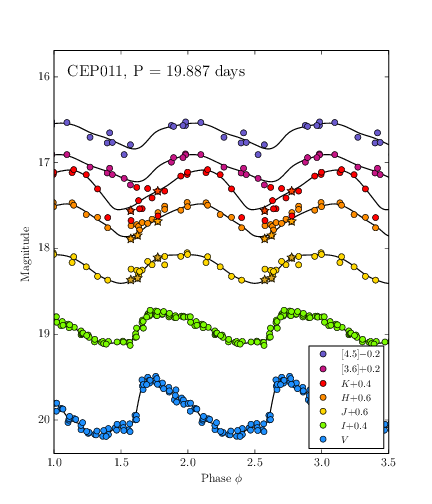}
\includegraphics[width=0.45\textwidth]{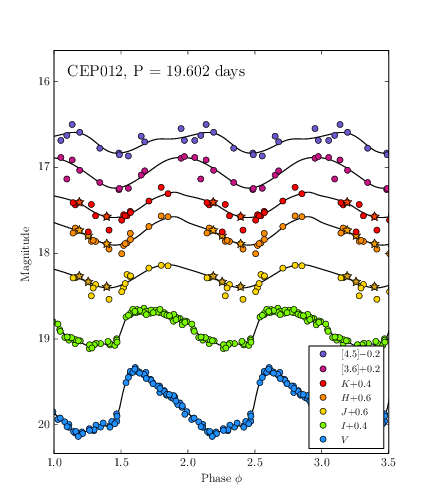}
\caption{Same as Figure A2, but for CEP009, CEP010, CEP011 and CEP012.}
\end{figure}

\begin{figure}
\centering
\includegraphics[width=0.45\textwidth]{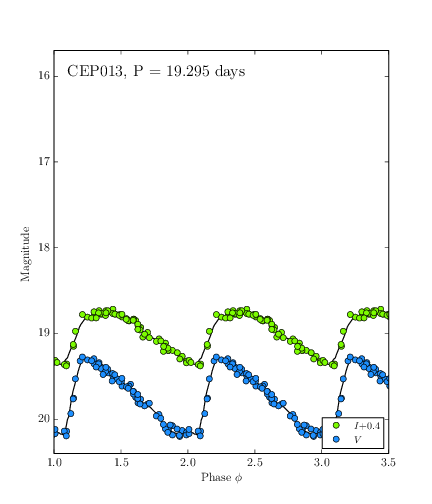}
\includegraphics[width=0.45\textwidth]{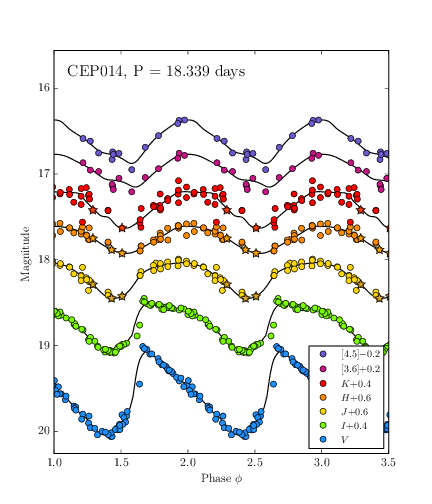}
\includegraphics[width=0.45\textwidth]{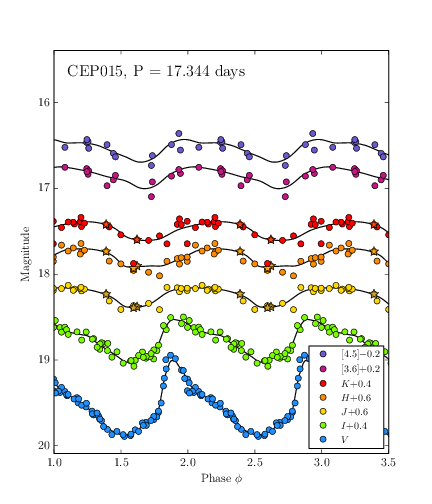}
\includegraphics[width=0.45\textwidth]{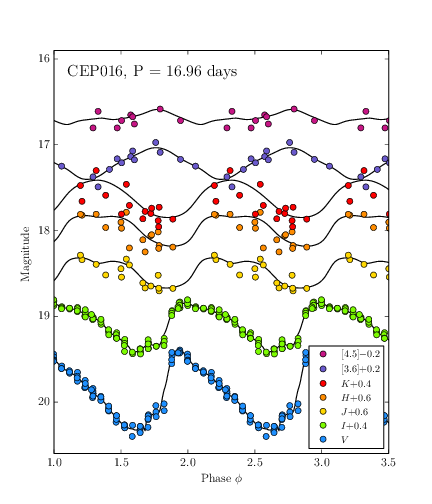}
\caption{Same as Figure A2, but for CEP013, CEP014, CEP015 and CEP016.}
\end{figure}

\begin{figure}
\centering
\includegraphics[width=0.45\textwidth]{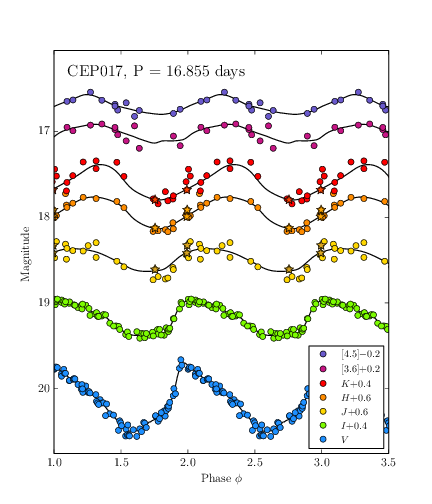}
\includegraphics[width=0.45\textwidth]{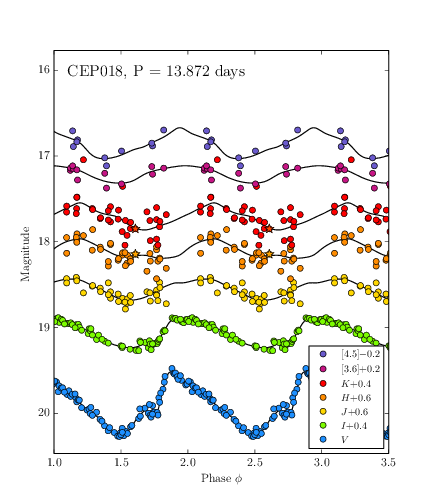}
\includegraphics[width=0.45\textwidth]{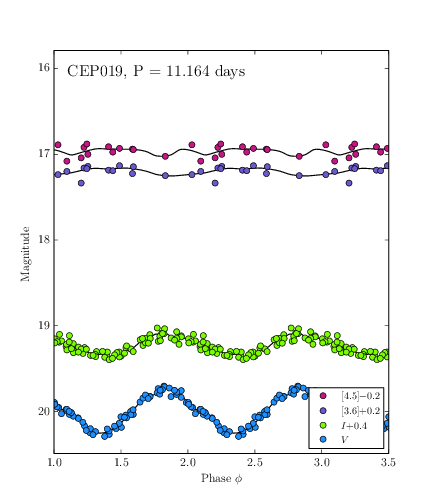}
\includegraphics[width=0.45\textwidth]{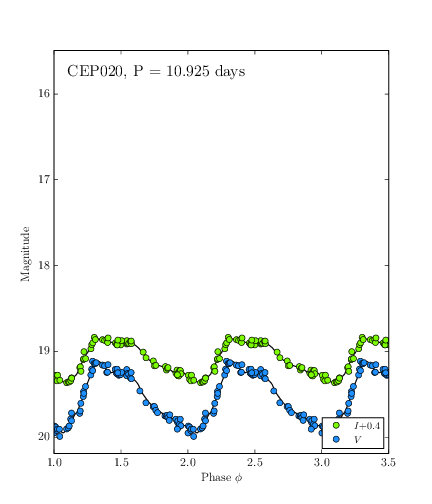}
\caption{Same as Figure A2, but for CEP017, CEP018, CEP019 and CEP020.}
\end{figure}

\begin{figure}
\centering
\includegraphics[width=0.45\textwidth]{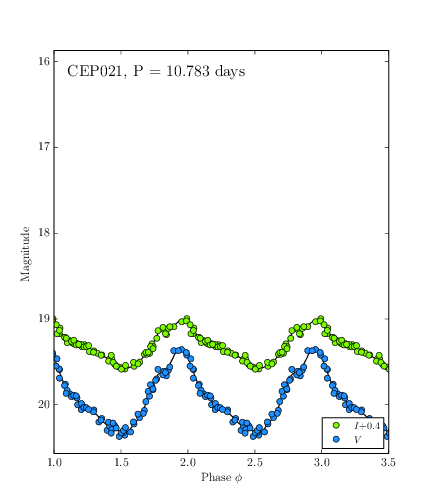}
\includegraphics[width=0.45\textwidth]{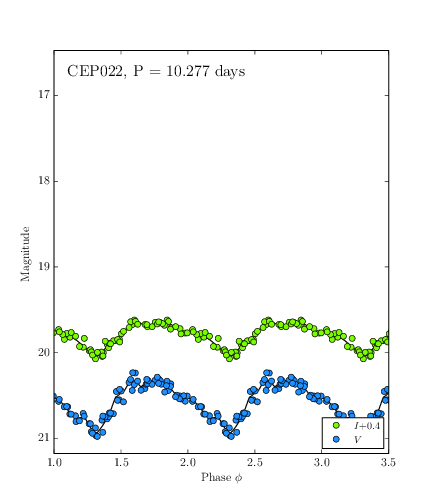}
\includegraphics[width=0.45\textwidth]{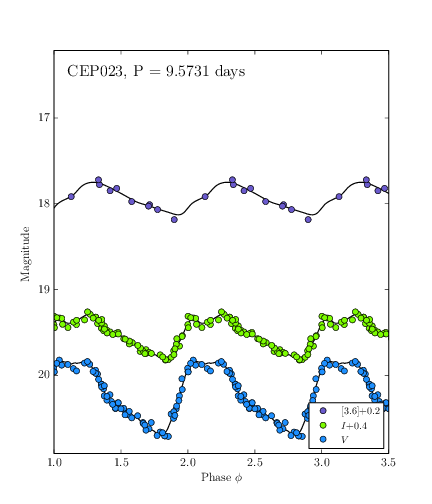}
\includegraphics[width=0.45\textwidth]{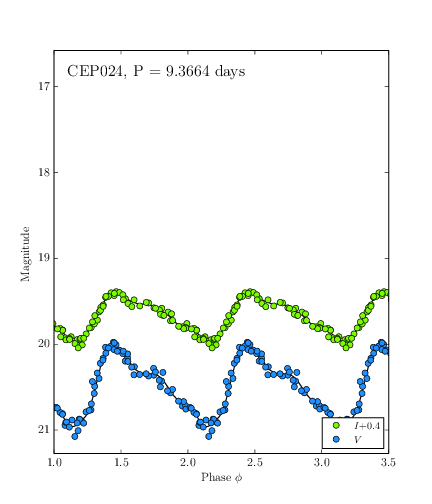}
\caption{Same as Figure A2, but for CEP021, CEP022, CEP023 and CEP024.}
\end{figure}

\begin{figure}
\centering
\includegraphics[width=0.45\textwidth]{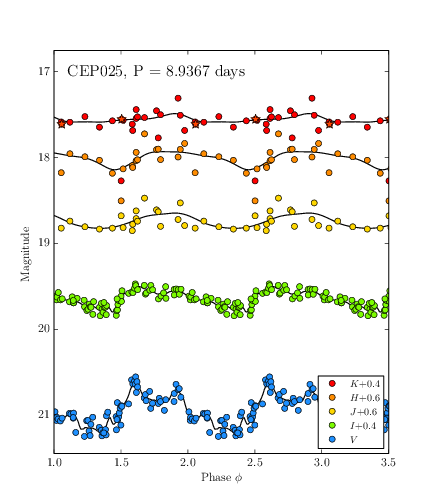}
\includegraphics[width=0.45\textwidth]{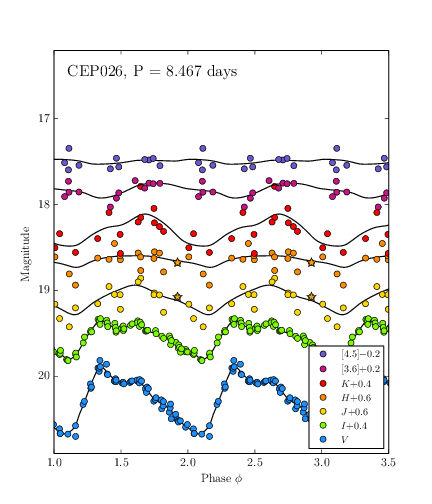}
\includegraphics[width=0.45\textwidth]{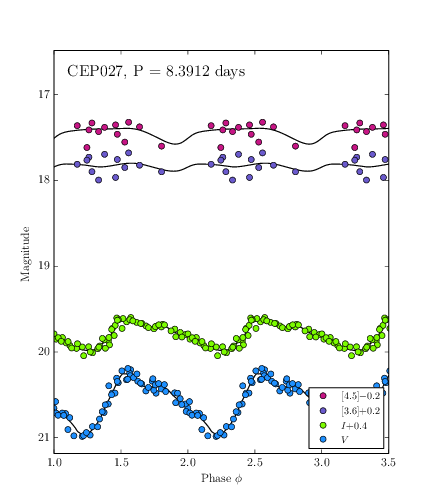}
\includegraphics[width=0.45\textwidth]{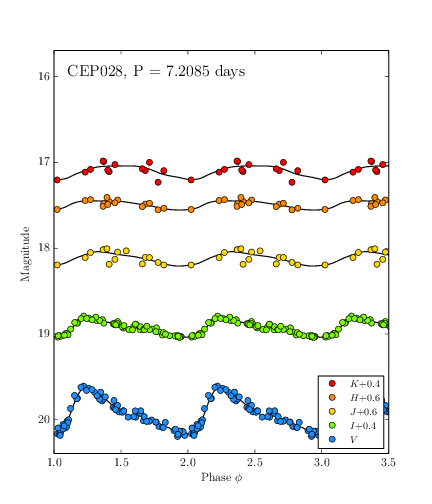}
\caption{Same as Figure A2, but for CEP025, CEP026, CEP027 and CEP028.}
\end{figure}

\begin{figure}
\centering
\includegraphics[width=0.45\textwidth]{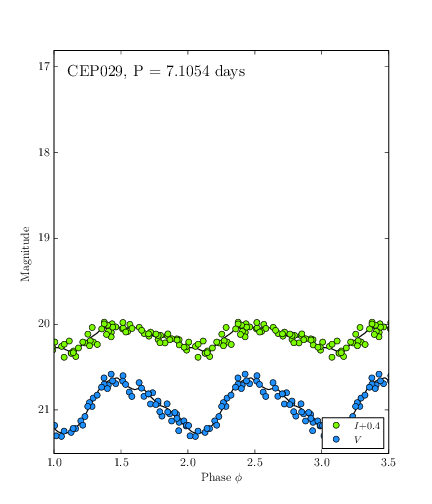}
\includegraphics[width=0.45\textwidth]{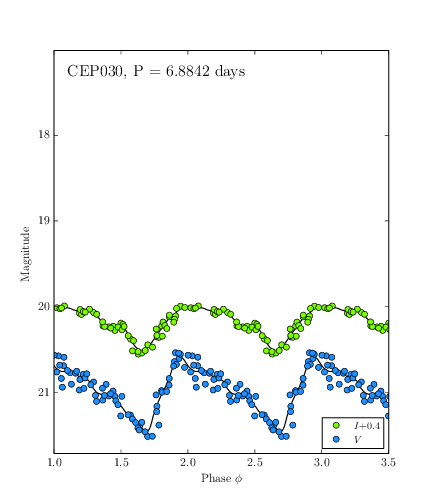}
\includegraphics[width=0.45\textwidth]{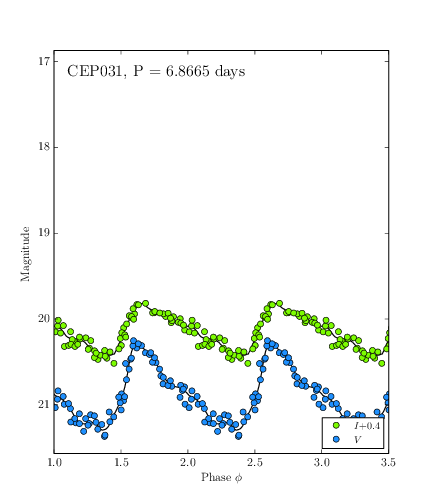}
\includegraphics[width=0.45\textwidth]{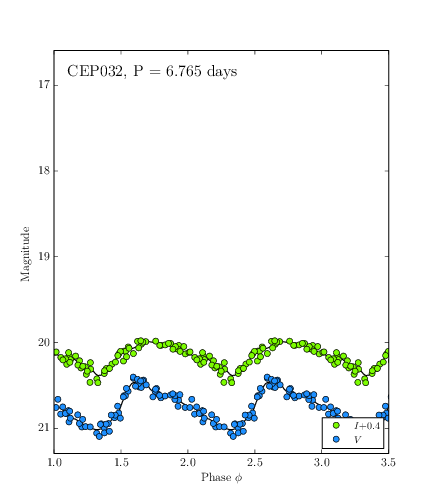}
\caption{Same as Figure A2, but for CEP029, CEP030, CEP031 and CEP032.}
\end{figure}

\begin{figure}
\centering
\includegraphics[width=0.45\textwidth]{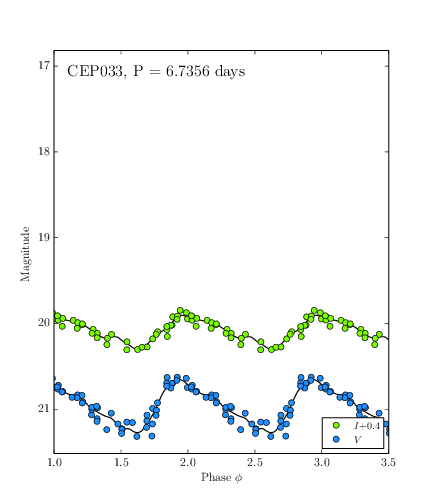}
\includegraphics[width=0.45\textwidth]{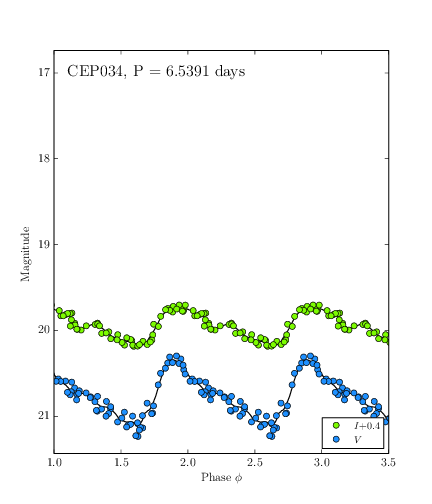}
\includegraphics[width=0.45\textwidth]{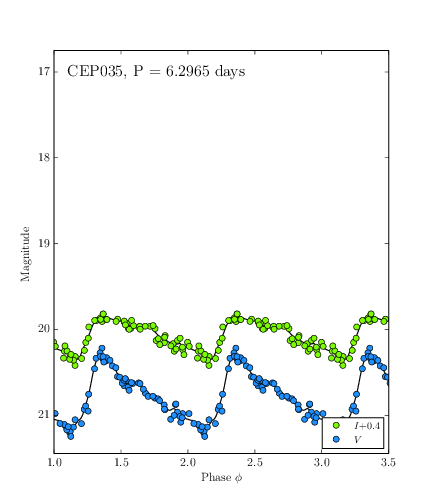}
\includegraphics[width=0.45\textwidth]{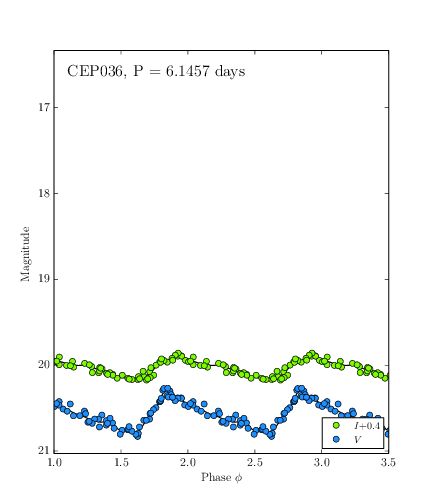}
\caption{Same as Figure A2, but for CEP033, CEP034, CEP035 and CEP036.}
\end{figure}

\begin{figure}
\centering
\includegraphics[width=0.45\textwidth]{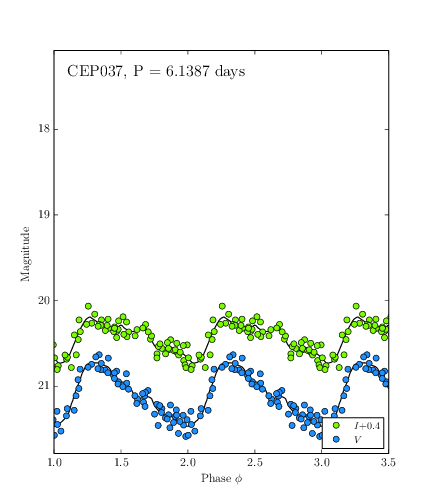}
\includegraphics[width=0.45\textwidth]{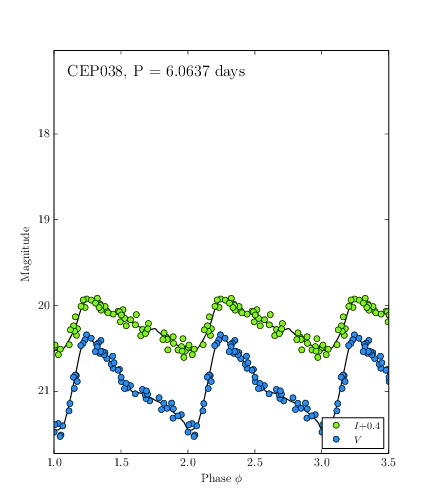}
\includegraphics[width=0.45\textwidth]{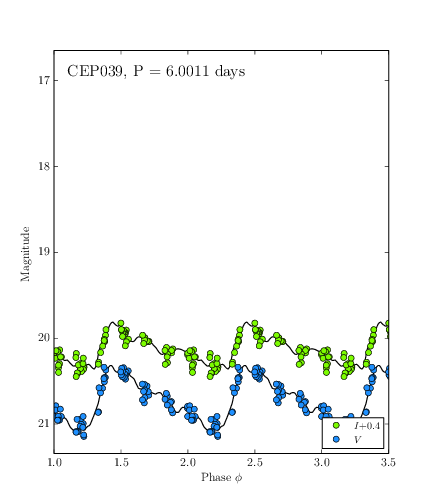}
\caption{Same as Figure A2, but for CEP037, CEP038 and CEP039.}
\end{figure}
\end{document}